\documentclass{INTERSPEECH2023}

\interspeechcameraready 
\usepackage{anyfontsize}
\usepackage{amsmath,graphicx}
\usepackage{hyperref}
\usepackage{amssymb}
\usepackage{url}
\usepackage{multirow}
\usepackage{paralist}
\usepackage{array}

\def\attn{\mathrm{attn}}

\def\RR{{\mathbb R}}

\newcolumntype{P}[1]{>{\centering\arraybackslash}p{#1}}

\title{Topological Data Analysis for Speech Processing}
\name{\begin{tabular}{cccc}
Eduard Tulchinskii$^1$, & Kristian Kuznetsov$^1$, & Laida Kushnareva, & Daniil Cherniavskii$^3$, \\
Serguei Barannikov$^{1,2}$, & Irina Piontkovskaya, & Sergey Nikolenko$^{4}$, & Evgeny Burnaev$^{1,3}$
\end{tabular}}

\address{\textsuperscript{1}Skoltech, Russia; \textsuperscript{2}CNRS, Université Paris Cité, France; 
\\\textsuperscript{3}Artificial Intelligence Research Institute (AIRI), Russia;
\\\textsuperscript{4}St. Petersburg Department of the Steklov Institute of Mathematics, Russia
}

\email{Eduard.Tulchinskiy@skoltech.ru}

\begin{document}
%
\maketitle

\begin{abstract}
We apply topological data analysis (TDA) to speech classification problems and to the introspection of a pretrained speech model, HuBERT. To this end, we introduce a number of topological and algebraic features derived from Transformer attention maps and embeddings. We show that a simple linear classifier built on top of such features outperforms a fine-tuned classification head. We 
achieve an improvement of about 9\% accuracy and 5\% ERR on two common datasets; on \emph{CREMA-D}, the proposed feature set reaches a new state of the art performance with accuracy 80.155. We also show that topological features are able to reveal functional roles of speech Transformer heads; e.g., we find the heads capable to distinguish between pairs of sample sources (natural/synthetic) or voices without any downstream fine-tuning. Our results demonstrate that TDA is a promising new approach for speech analysis, especially for tasks that require structural prediction.
\end{abstract}



\section{Introduction}

The paradigm of learning universal large-scale Transformer-based models has been recently transferred to speech from natural language processing~\cite{9585401,baevski2020wav2vec,chung2021w2v}. 
Due to the high complexity of speech data, the preferred way of downstream task adaptation for such models is to keep pretrained weights frozen and fine-tune small task-specific heads~\cite{Yang2021SUPERBSP}, so it is important to use the model's hidden states in the most efficient way. A common approach is to use the outputs of the last layer of the model, combining them via various pooling methods, although it has been shown that for some tasks, e.g. phoneme prediction in HuBERT or the analogy task in BERT, embeddings from lower and middle layers are more useful~\cite{vulic2020probing}. \emph{Topological data analysis} (TDA) is a recently proposed way to obtain more efficient data representations from frozen Transformer weights~\cite{kushnareva-etal-2021-artificial,judgements}. TDA features prove to be better suited for many downstream tasks, including artificial text detection and linguistic acceptability judgement. In particular, for ungrammatical sentence detection conventional sentence embeddings yield classification quality no better than random, while TDA has led to meaningful results. Inspired by these results, in this work we apply TDA to the HuBERT model to construct more powerful speech representations.

TDA has already been applied to signals of various nature. Previous attempts for speech used topological properties of the audio signal: persistent entropy for noise classification~\cite{RUCCO2017130} and emotion recognition~\cite{Gonzalez-Diaz}, detection of a periodic signal in noisy data~\cite{9747228} etc. However, TDA for Transformer-based models has so far been limited to natural language processing (NLP): TDA for attention maps for artificial text detection~\cite{kushnareva-etal-2021-artificial}
and linguistic acceptability~\cite{judgements}, TDA for word embeddings for dialogue term extraction~\cite{vukovic-etal-2022-dialogue} and constructing story trees~\cite{haghighatkhah-etal-2022-story}. The evolution of inner representations in neural networks has been studied with persistent Betti numbers~\cite{topologyofdeep} and recently introduced representation topology divergence (RTD)~\cite{barannikov2021representation}, which we apply in this work
to intermediate embeddings and attention maps of a speech Transformer.

In this work we also apply TDA to interpreting pretrained Transformer models. It is well known in NLP that different heads are sensitive to different phenomena~\cite{vulic2020probing}, and we demonstrate the same effect for speech Transformers, finding heads that are best for solving specific downstream tasks: separating a given pair of emotions, a given pair of speakers, detecting speech generated by a specific TTS model, or representing spectral features of sound samples (bit rate, LPCC etc.). 
Below, Section~\ref{sec:methods} introduces TDA, Section~\ref{sec:results} shows our evaluation study, Section~\ref{sec:interpretation} applies TDA to interpreting attention maps, and Section~\ref{sec:concl} concludes the paper.

\section{Methods}\label{sec:methods}

Below, we first introduce topological data analysis for weighted graphs and then define the features we extract from HuBERT.

\noindent
\textbf{Homologies of graphs and manifolds}. 
Topology is the study of properties of manifolds in space other than their size, i.e., 
loosely speaking, topology considers an object as if it were made of an infinitely stretchable material that can
stretch and compress as long as there are no cuts or gluings.
\textit{Topological invariants} are properties of manifolds that do not change under topological transformations. \textit{Homology groups} are among the most important invariants, calculated separately for different dimensions. The homology group of dimension $i$ consists of $i$-dimensional closed objects on the manifold that cannot be converted into each other by a topological transformation; e.g., the $0$-th homology group consists of $0$-dimensional objects (points) and equals in size to the number of connected components. In this work, we are not going beyond the $0$-th dimension, but our methods can easily be extended to the $H_1$ group, which consists of the closed paths on the manifold, or cycles in a graph. 
For higher dimensions, the computations may become prohibitive.

\begin{figure}[!t]
    \includegraphics[width=\linewidth]{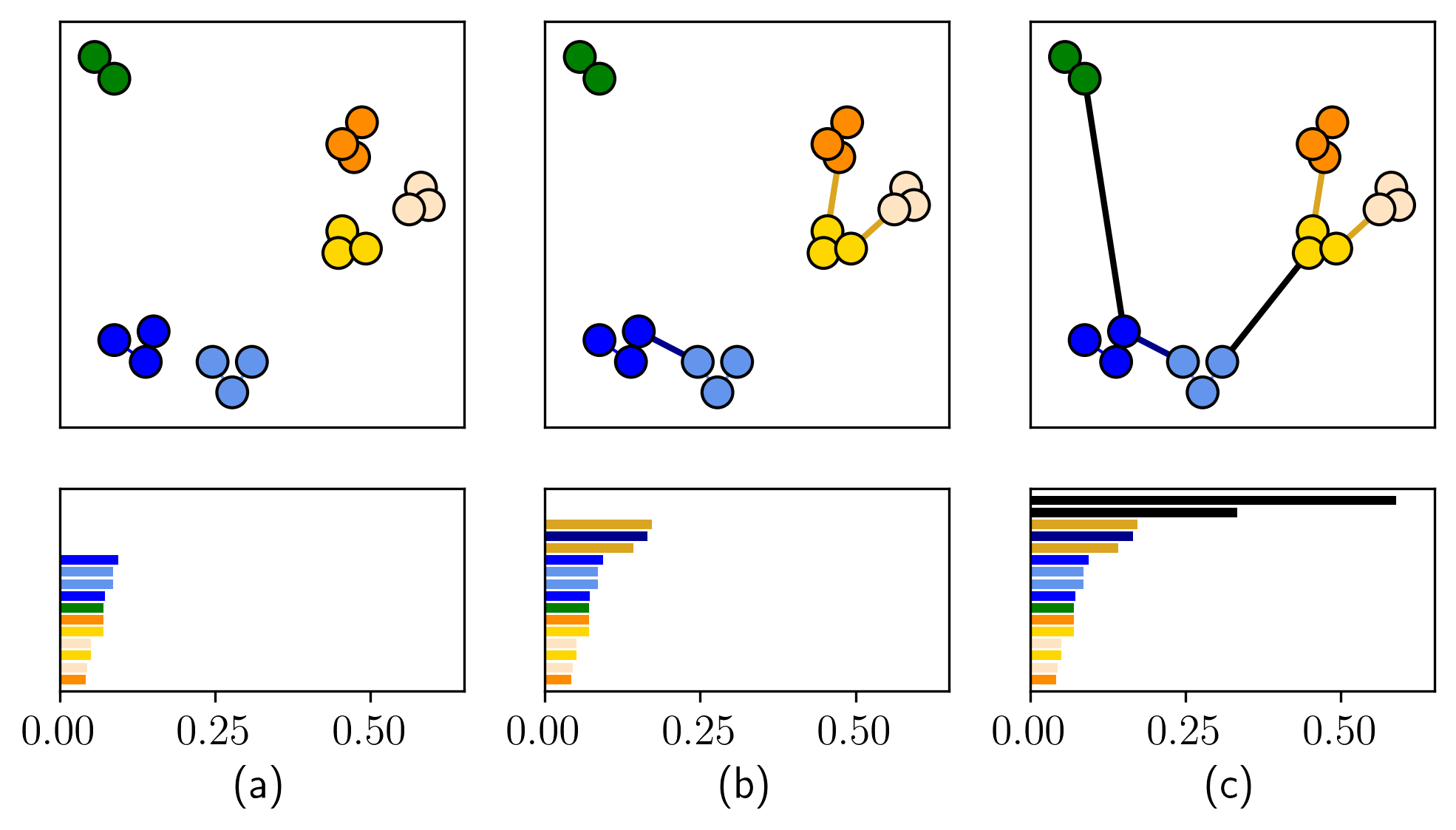}
\caption{Constructing an $H_0$ barcode for a dataset with a hierarchical structure: \textit{(a)} first, small clusters are connected; \textit{(b)} as the threshold grows, larger blue and yellow clusters are joined; (c) finally, all dots are joined into a single component. Each bar corresponds to an edge in the minimal spanning tree.}\label{fig:barcodes}\vspace{-.2cm}
\end{figure}

\noindent
\textbf{Persistent homologies for weighted graphs and point clouds}. 
We would like to apply TDA to sets of vectors in $\RR^d$ (point cloud). This set can be viewed as a complete graph $G$ with edges weighted by a distance-like metric between vectors. To estimate the topological properties of this graph, it looks natural to remove weak connections by thresholding, leaving only edges with weights lower than a given $\epsilon$. However, it is not clear which threshold to choose. TDA can track the changes of topology across varying thresholds via \emph{persistent homologies}. 
Fig. \ref{fig:barcodes} illustrates the persistence of $0$-dimensional homologies $H_0$ (connected components) for a set of points. For $\epsilon$ below the minimal distance between vertices, we obtain a graph with no edges; as $\epsilon$ increases, new edges are added, ending in the complete graph. During this process, gradual changes of graph topology can be expressed in terms of the ``birth'' and ``death'' of basic features. We begin with $|V|$ connected components (all of them are ``born''), and as $\epsilon$ increases, pairs of them are joined together (one component ``dies''). ``Birth'' and ``death'' moments can be represented with a diagram called the \emph{barcode}~\cite{barannikov2021canonical,10.3389/frai.2021.667963}, where the horizontal axis is a sequence of thresholds $\epsilon$, and each horizontal bar corresponds to a single feature. 

An important property of the $H_0$ barcode is that its bars correspond to the edges of the minimal spanning tree (MST) of the graph. Indeed, our process of adding edges as the threshold grows coincides with the classical Prim algorithm that finds the MST in a weighted graph; the length of the bar which is ``dying'' on every step equals the length of the corresponding MST edge. We denote by $H_0^m$ the average length of bars in the $H_0$ barcode.

The $H_0$ barcode can capture information about the hierarchical structure of the dataset,  as illustrated in Fig.~\ref{fig:barcodes}. As we show in Section~\ref{sec:interpretation}, our methods work well with the hierarchical information of speech signals, distinguishing the levels of frames, phonemes, words, and pauses. 

\emph{Representation topology divergence} (RTD) measures the topological dissimilarity between two data representations, formalized as two weighted graphs $G^a$ and $G^b$ with a common set of vertices. If RTD equals zero then the $H_0$ barcodes of the two graphs coincide. And vice versa, if the barcodes coincide and the resulting clusters of vertices are the same on all levels then RTD is zero. For details and the formal definition, see~\cite{barannikov2021representation,barannikov1994}.

\noindent
\textbf{Features}. 
We use three groups of features: algebraic features of attention matrices, topological features of attention matrices, and topological features of embeddings; for comparison, we also pool embeddings from all layers.
%
We consider 
HuBERT-base model~\cite{9585401} that consists of $12$ layers with $12$ attention heads each, with embedding dimension $768$.

\def\hzsym{H_0^{m,\mathrm{sym}}}
\def\hzpc{H_0^{m,\mathrm{pc}}}

\emph{Algebraic features} include the sum of the upper triangular part of the $n \times n$ attention matrix (normalized by $n^2$), which is used as a measure of asymmetry, and mean values of its $3$ longest diagonals. This yields $4$ features per attention map, $576$ for the entire HuBERT model.

\emph{Topological features of attention matrices} include the $H_0^m$ feature for two graphs derived from each attention matrix $A_{\attn}$. $\hzsym$ is defined as $H_0^{m}$ for the graph with adjacency matrix 
\begin{align}\label{eq:adjuc}
A^\prime = 1 - \max\left(A_{\attn},  A_{\attn}^\top\right)
,
\end{align}
that is, the symmetrization of $A_{\attn}$ (cf.~\cite{kushnareva-etal-2021-artificial}). $\hzpc$ is defined as $H_0^{m}$ for the rows of $A_{\attn}$ considered as a point cloud with $L_1$-distance. The intuition behind this choice of distance is that these points always lie on the $L_1$ sphere because the rows of attention matrix are normalized by \emph{softmax}. Here we have two features per attention map, $288$ in total.

\emph{Topological features of embeddings}. Considering the $i$-th layer's embeddings $X^{(i)}$ as a point cloud with the $L_2$-distance, we obtain $3$ features for each layer: $H_0^m(X^{(i)})$, RTD between $X^{(i)}$ and the last layer's embeddings $X^{(L)}$, and RTD between $X^{(i)}$ and initial embeddings $X^{(0)}$ (36 features in total). Besides, we add mean mel-frequency coefficients (MFCC); since we use 13 MFCCs, it gives 13 more features. Finally, we add two features: $H_0^m$ of the MFCC features and of the initial embeddings. In total, we have $51$ features for the entire model.

\emph{Embeddings from all layers}.
An alternative approach, inspired by~\cite{kenton2019bert}, is to use pooled embeddings from each layer of the Transformer, not just the last. We explored two pooling strategies: averaging over the timescale (common for speech Transformers) and taking only the first embedding as suggested in~\cite{kenton2019bert}. This yields $9216$ features in each case.

\section{Evaluation}\label{sec:results}

\noindent
\textbf{Datasets}.
We evaluate our method on tasks of emotion recogntion, antispoofing, and speaker verification. Our features are not applicable directly to ASR tasks because they characterize a speech sample as a whole; however, we additionally performed an evaluation on the word classification task to check that linguistic information can be captured. 
We have used standard datasets for these tasks. \textbf{IEMOCAP}, introduced in~\cite{Busso2008IEMOCAPIE}, contains $\approx 12$ hours of audiovisual data, including video, speech, motion capture of faces, and text transcriptions; we use only speech samples from the ``Anger'', ``Sadness'', ``Happiness'', and ``Neutral'' classes ($4490$ samples), with 5-fold cross-validation, similar to~\cite{8639633,9747460}.
\textbf{CREMA-D}~\cite{6849440}
has $7442$ clips from $91$ actors who spoke $12$ different sentences in one of six basic emotions (anger, disgust, fear, happy, neutral, and sad); we perform multi-class classification with $6$ classes 
and evaluate by averaging five splits with 70/15/15\% train/development/evaluation subsets.
\textbf{ASVSpoof}~\cite{9358099} was presented for the 3rd Automatic Speaker Verification Spoofing and Countermeasures Challenge; we used the standard split with $25380$ training, $24844$ development, and $71237$ evaluation samples and
performed classification into generated and bonafide labels, a standard task on ASVSpoof~\cite{Yu2018SpoofingDI}.
\textbf{VoxCeleb1}~\cite{nagrani17_interspeech} contains over $100$K utterances by $1251$ celebrities extracted from \emph{YouTube} videos; we performed binary classification between pairs of utterances from the same or different speakers, with $40000$ pairs in the training set, $8000$ in development, and $37720$ in the test set; in this dataset, we clip each utterance to the first $5$ seconds for all methods, due to the amount of data.  
\textbf{FSDD}\footnote{{\scriptsize \url{https://doi.org/10.5281/zenodo.1342401}}} (Free Spoken Digit Dataset) is a simple dataset consisting of $3000$ short recordings of spoken digits. Each recording was obtained from one of $6$ speakers, and there are 50 recordings of each digit per speaker. The test set consists of 10\% of the recordings that include all speakers and digits. We performed multi-class classification on this task.


\noindent
\textbf{Models}.
We utilize the \emph{HuBERT Base} model~\cite{9585401} pretrained for automatic speech recognition on 960 hours of audio from the \emph{Librespeech} corpus~\cite{7178964}. We use HuBERT as a pretrained frozen instance, without fine-tuning or any other adjustment of the weights.
We use a linear layer trained over the pooled output of the Transformer as the baseline, which is consistent with the SUPERB leaderboard~\cite{Yang2021SUPERBSP}; we used the results for IEMOCAP and VoxCeleb1 published there and trained the baselines ourselves for the other datasets.

As our TDA-based approach, we train a logistic regression model with $L_1$-regularization over the set of features computed from HuBERT attention maps and/or embeddings. For automatic speaker verification (checking if two utterances are made by the same person) we compute the absolute value of elementwise differences between features of both utterances. 



\begin{table}[!t]\centering\small
\setlength{\tabcolsep}{1pt}

\caption{Experimental results; $\star$~--- from SUPERB~\cite{Yang2021SUPERBSP}.}
\label{tbl:results}
\begin{tabular}
{P{.24\linewidth}P{.15\linewidth}P{.15\linewidth}P{.13\linewidth}P{.13\linewidth}P{.12\linewidth}}\hline
 \textbf{Model} & \textbf{IEMO CAP} & \textbf{CREMA-D} & \textbf{ASV Spoof} & \textbf{Vox Celeb1} & \textbf{FSDD} \\
 & Acc $\uparrow$ & Acc $\uparrow$ & {\footnotesize EER} $\downarrow$ & {\footnotesize EER} $\downarrow$ & Acc $\uparrow$ \\ \hline
HuBERT (baseline) & $64.92^\star$ & $71.047$ {\footnotesize $\pm0.566$} & $6.649$ & \hspace{0.1cm} $\boldsymbol{7.45}$ & $\underline{99.3}$ {\footnotesize $\pm1.3$}\\\hline
All layer embs, $1$st & $65.612$ {\footnotesize $\pm1.050$} & $71.320$ {\footnotesize $\pm0.479$} & $2.706$ & $46.240$ & $96.0$ {\footnotesize $\pm0.7$} \\\hline
All layer embs, mean  & $69.355$ {\footnotesize $\pm1.801$} & $76.260$ {\footnotesize $\pm1.148$} & $\boldsymbol{1.519}$ & \hspace{0.1cm} $\underline{8.46}$  
 & $97.7$ {\footnotesize $\pm0.5$} \\\hline
Attention features & $\underline{69.666}$ {\footnotesize $\pm1.174$} & $\underline{79.200}$ {\footnotesize $\pm1.240$} & $2.138$ & $30.326$ & $98.7$ {\footnotesize $\pm0.8$} \\\hline
Attn. \& non-attn. feat. & $\boldsymbol{69.955}$ {\footnotesize $\pm\boldsymbol{0.972}$} & $\boldsymbol{80.155}$ {\footnotesize $\pm\boldsymbol{0.680}$} & $\underline{1.946}$ & $26.443$ & $\boldsymbol{99.6}$ {\footnotesize $\pm\boldsymbol{0.4}$} \\\hline 
\end{tabular}
\end{table}

\noindent
\textbf{Experimental results}.
Table~\ref{tbl:results} shows experimental results on all five datasets; we report accuracy (Acc, in $\%$) and equal error rate (EER, in $\%$) for models trained on four different sets of features (see Section~\ref{sec:methods}):
\begin{inparaenum}[(1)]
\item \emph{Attention features} is a combination of {algebraic} and {topological features} calculated from attention maps;
\item \emph{Attn. \& non-attn. feat.} denotes a combination of algebraic and topological features of attention maps and topological features of embeddings; this combines all our TDA features for this task;
\item \emph{All layer embs, 1st} is the concatenation of first embeddings from all HuBERT layers;
\item \emph{All layer embs, mean} is the concatenation of all HuBERT embeddings with timescale averaging.
\end{inparaenum}
Below we show our conclusions from Table~\ref{tbl:results}.

First, on all datasets averaging the embeddings is a much better strategy than taking just the first. 
%
One reason might be that in HuBERT data representation there is no a fictional first token that represents the sample as a whole, similar to [CLS] in NLP Transformers.
Second, adding non-attention features improves performance compared to just {attention-based} features, as expected since they contain more information, although for IEMOCAP the improvement is quite marginal.
Third, topological features give better results than embeddings from all layers on multi-class emotion recognition datasets (IEMOCAP, CREMA-D).
%
On emotion recognition datasets (IEMOCAP and CREMA-D) we achieve major improvements over the baseline (conventional usage of HuBERT). Our results on VoxCeleb1 are very close to the baseline (the second best values in every column of Table~\ref{tbl:results} are underlined), and on FSDD our method performs slightly better than the baseline and very close to perfect accuracy. 
For CREMA-D, we achieve a new state of the art result, improving the previous SOTA of 70.47\%~\cite{ristea22_interspeech} by over 9\%. 

For emotion recognition tasks, topological features provide clear benefits over all considered baselines. For generated speech detection and speaker verification, attention features do not help with task solving in general, but in the next section we show that features computed for individual heads perform surprisingly well for restricted subtasks in the zero-shot setting.

\section{TDA for interpreting attention maps}\label{sec:interpretation}

\noindent
\textbf{Restricted tasks}.
Inspired by recent works on attention interpretability in natural language processing~\cite{vulic2020probing}, we analyze the roles of individual attention heads in HuBERT.
 Since their ``areas of expertise'' are much more narrow than general problems considered above, we use two restricted tasks: separation of individual models (one synthetic model vs real speech) and separation (binary classification) of two speakers.

\begin{figure}[!t]\centering
\setlength{\tabcolsep}{0pt}
\def\mywid{0.49\linewidth}
\begin{tabular}{cc}\centering
    \includegraphics[width=\mywid]{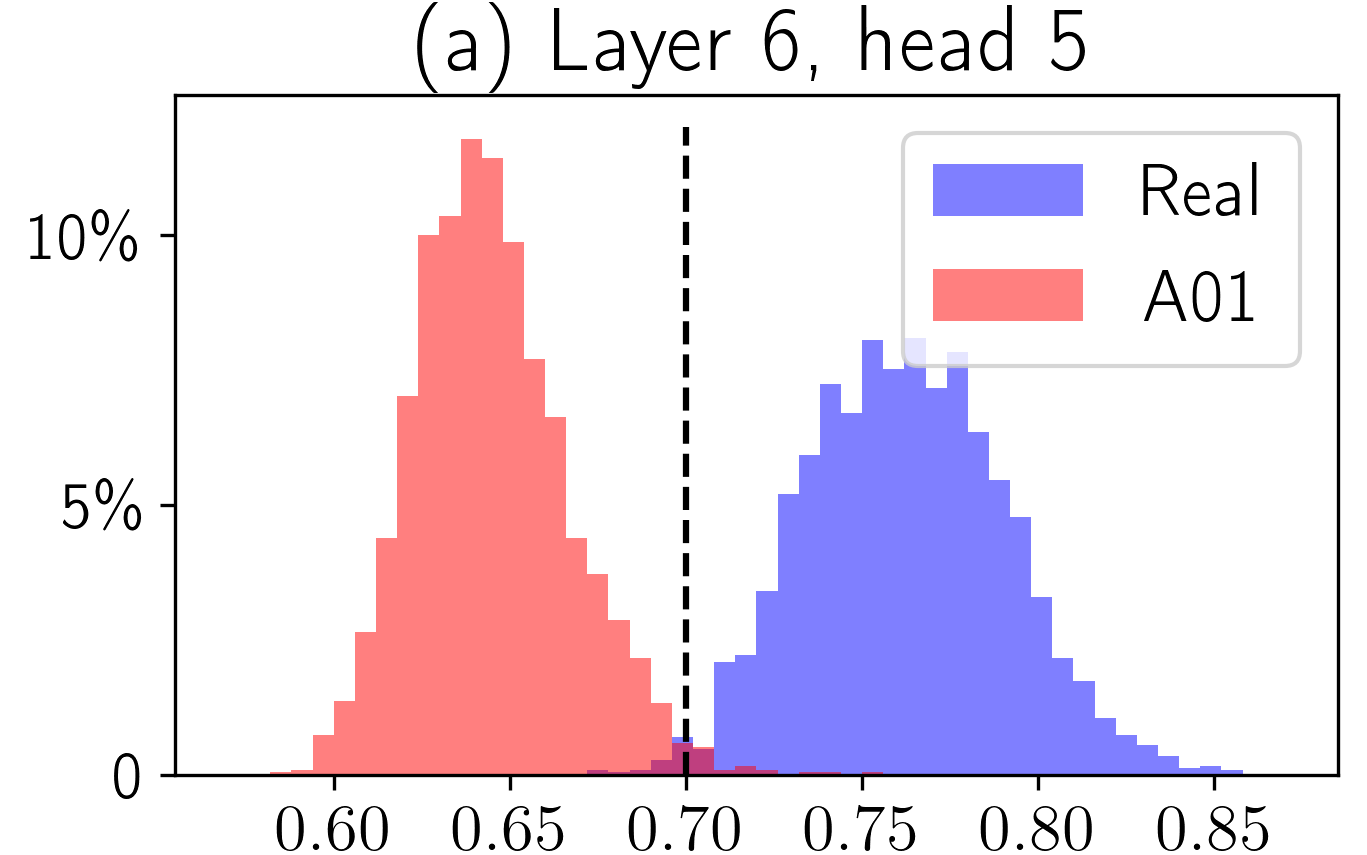} &
    \includegraphics[width=\mywid]{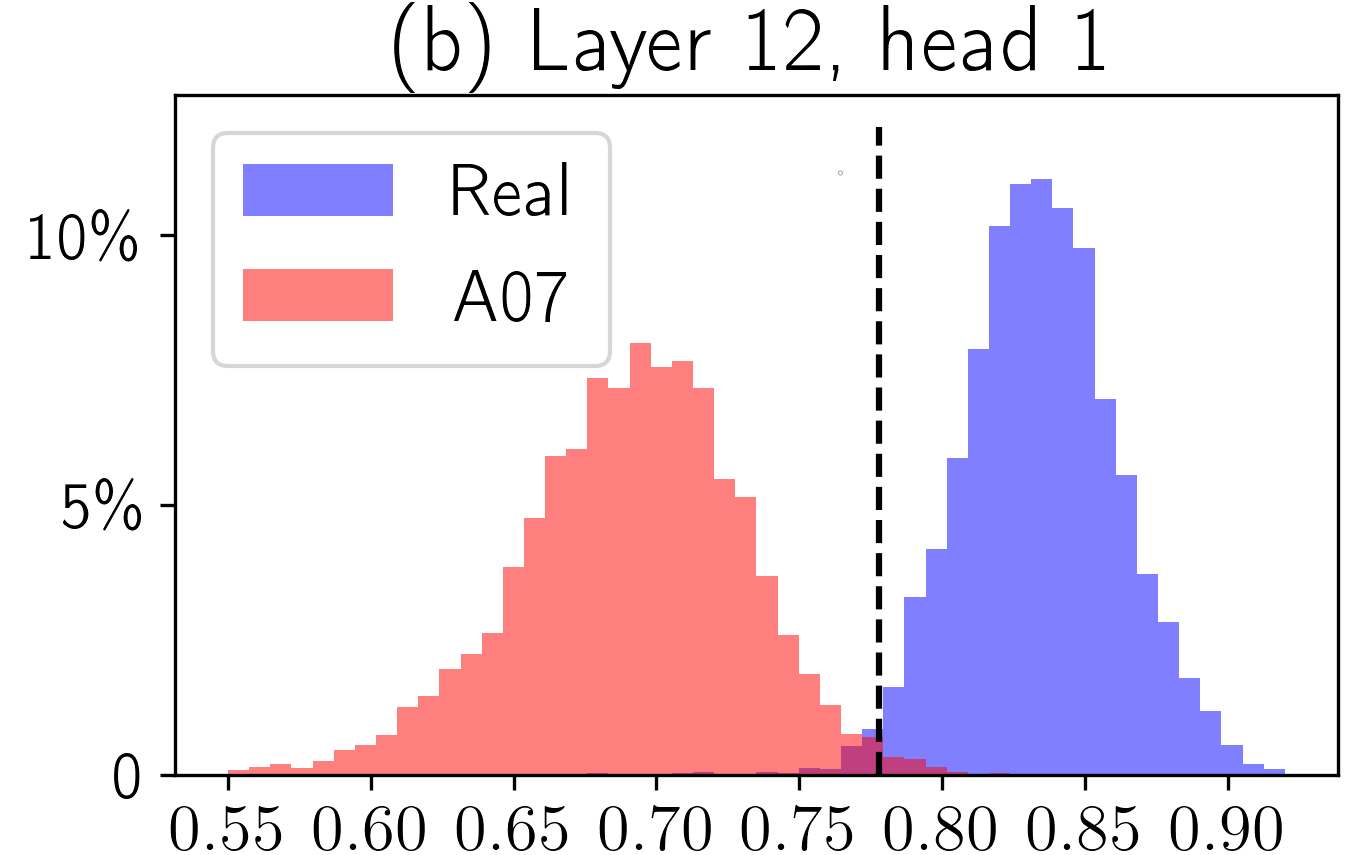} \\
    \includegraphics[width=\mywid]{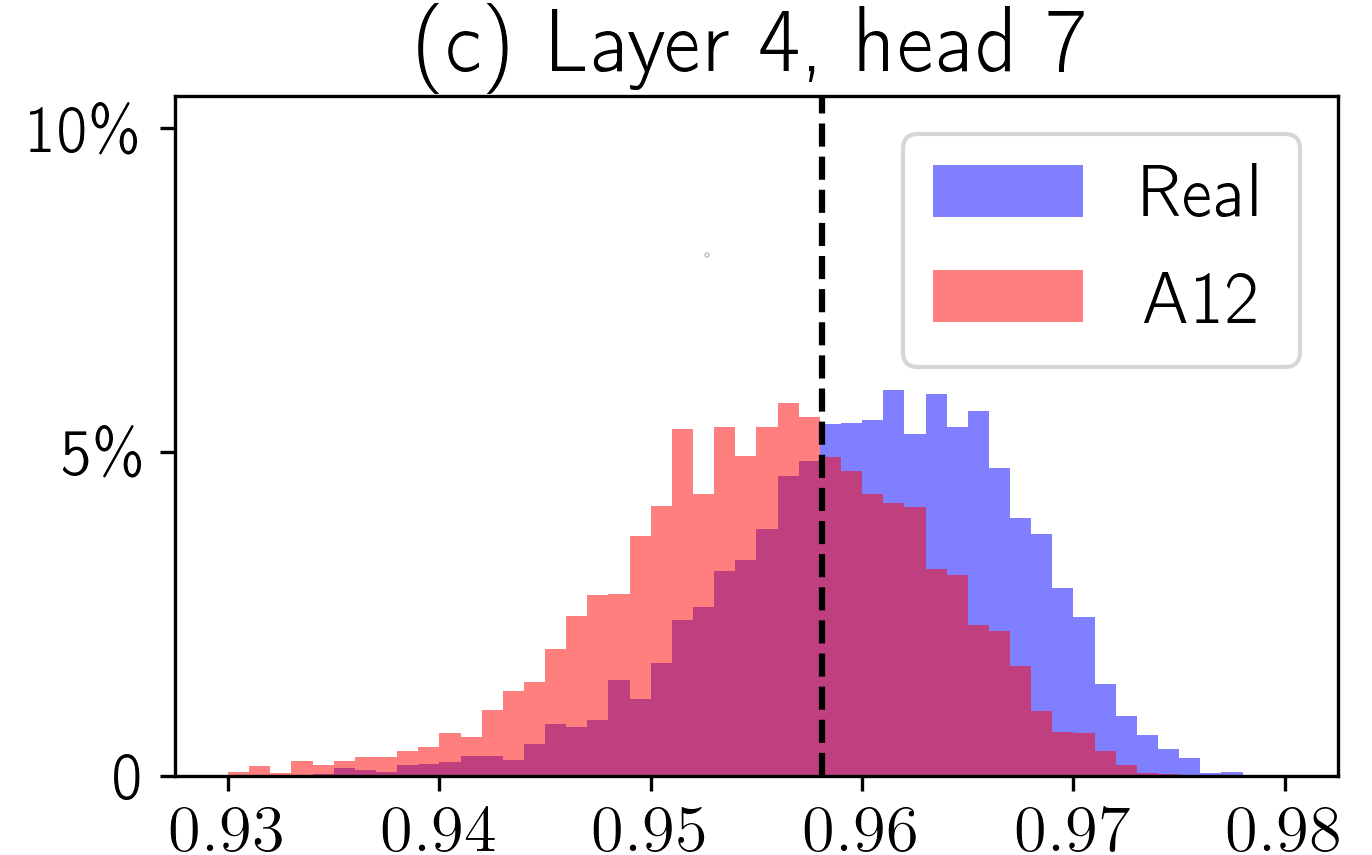}
    &
    \includegraphics[width=\mywid]{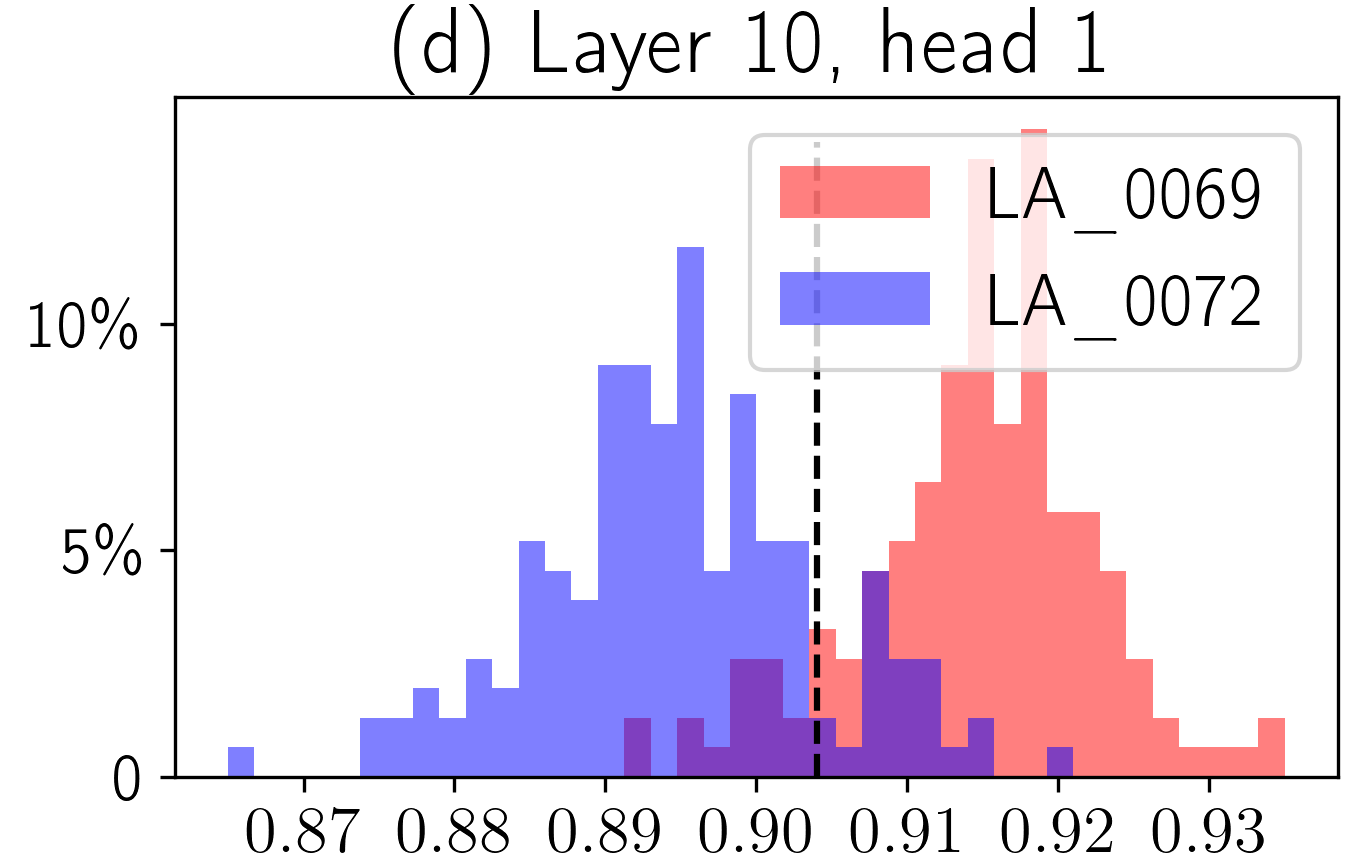} \\
    \includegraphics[width=\mywid]{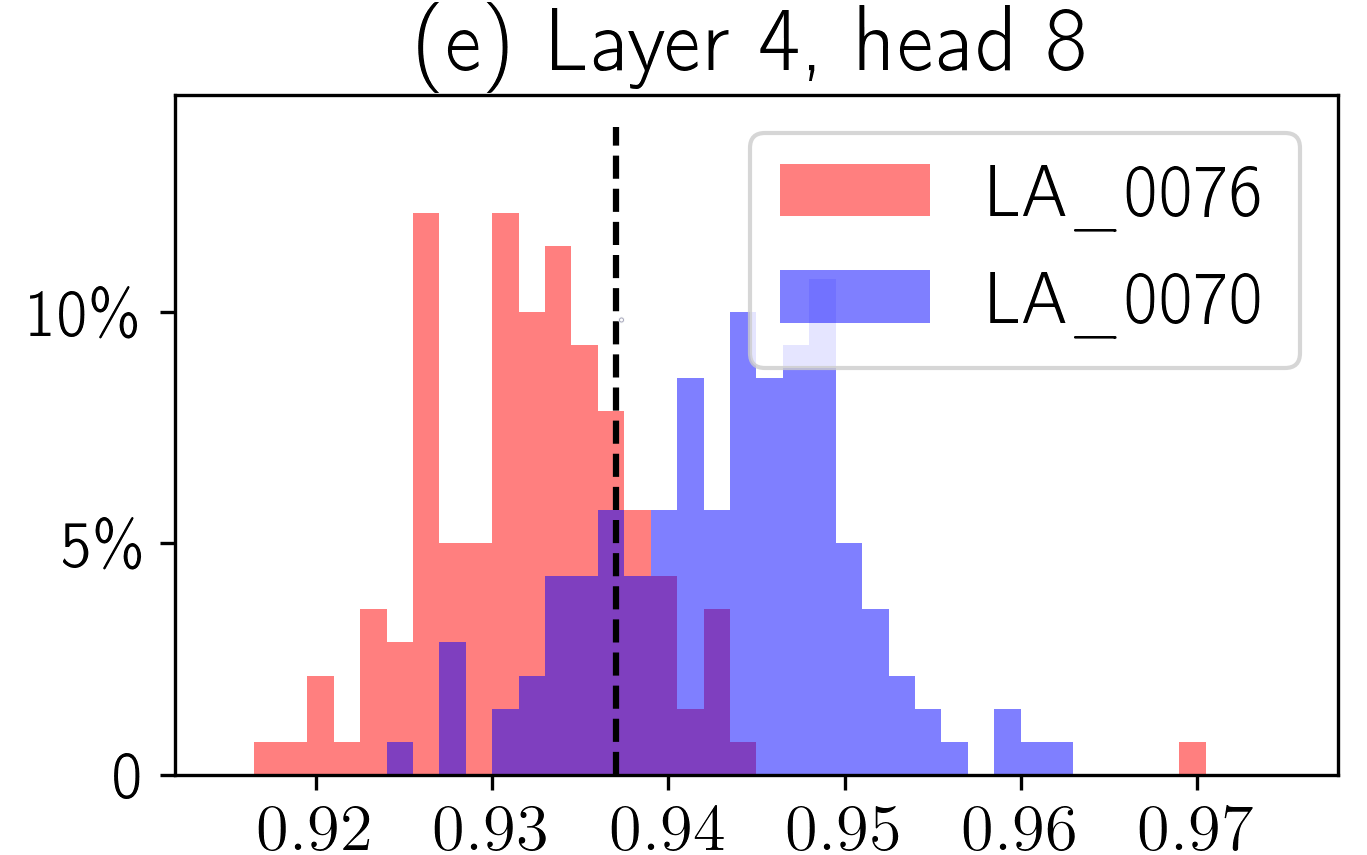} &
    \includegraphics[width=\mywid]{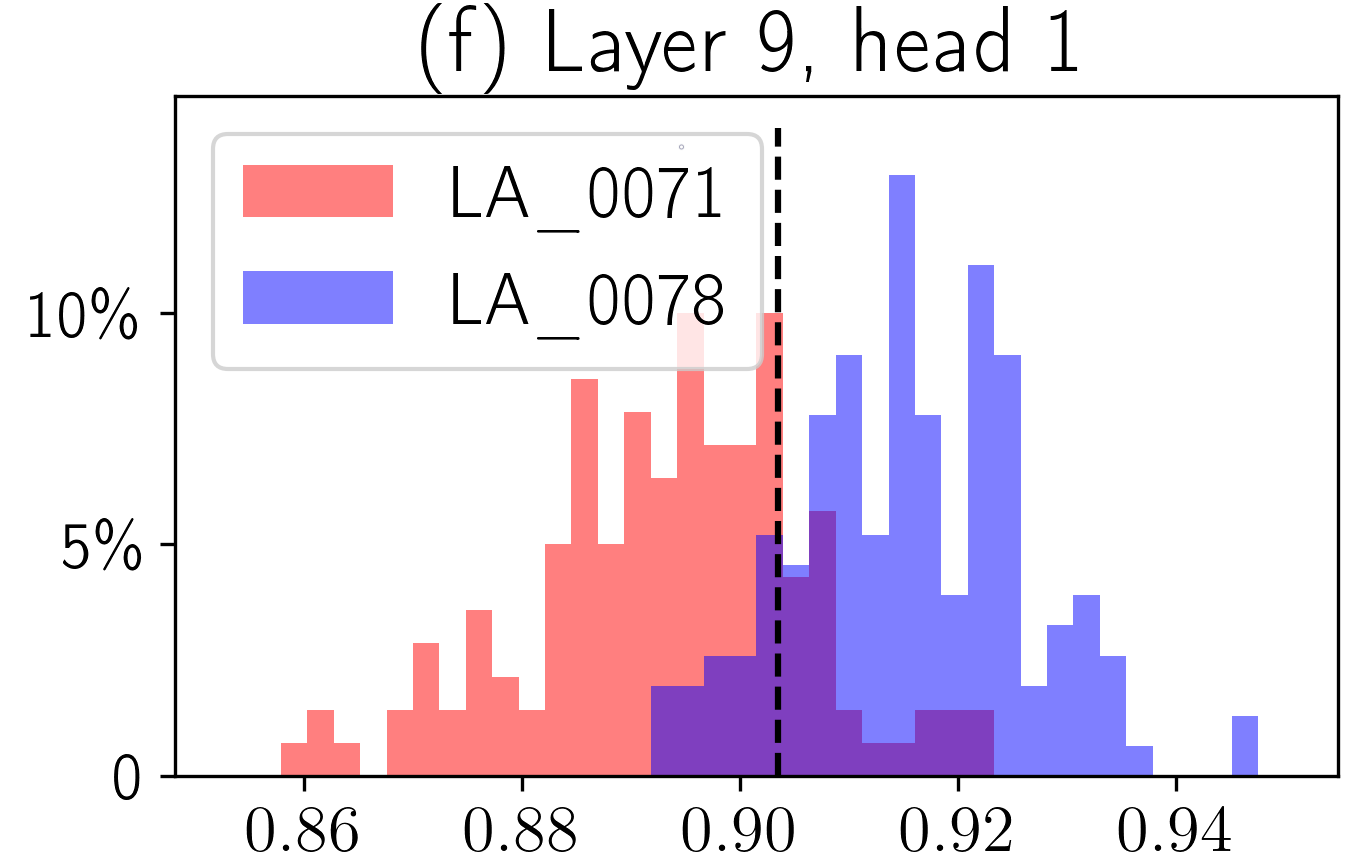}
\end{tabular}
    \caption{$\hzsym$ for the best HuBERT heads for two tasks: (a-c) individual model separation between human (blue) and synthetic (red) speech; (d-e) speaker separation; LA\_0069, LA\_0072, LA\_0078 -- female speakers, LA\_0070, LA\_0076, LA\_0071 -- male speakers.}\label{fig:sep_charts}    
\end{figure}

\begin{figure}[!t]\centering
    \includegraphics[width=.85\linewidth]{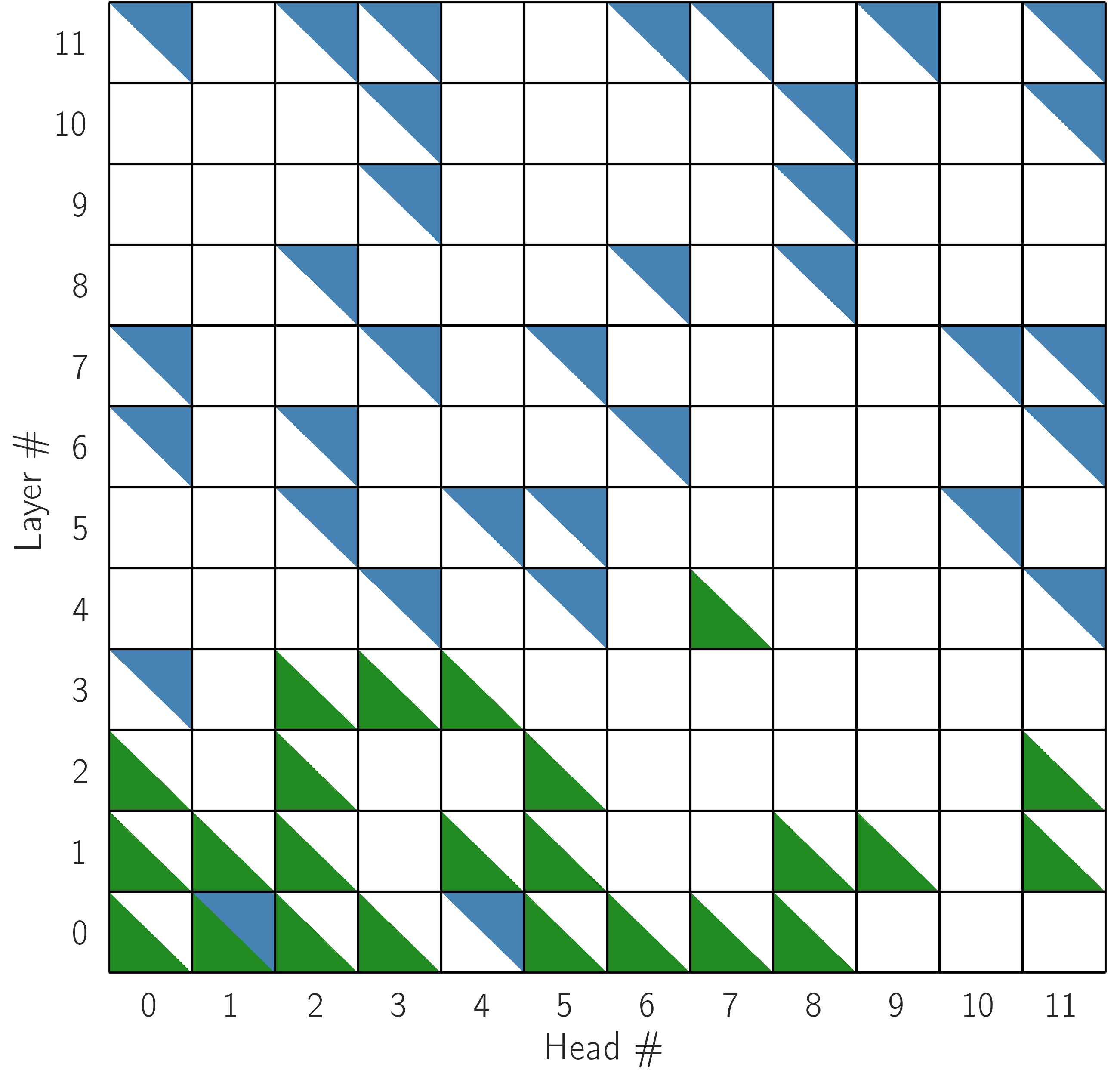}
    \vskip-.1in
\caption{Distribution of heads with high correlation (Pearson correlation coefficient $\geq 0.5$) of $\hzpc$ and standard acoustic features: blue, MFCC; green, PLP.}\label{fig:pcc_TDA_spectral}   \vskip-.1in
\end{figure}

\begin{figure}[!t]\centering
    \includegraphics[width=\linewidth]{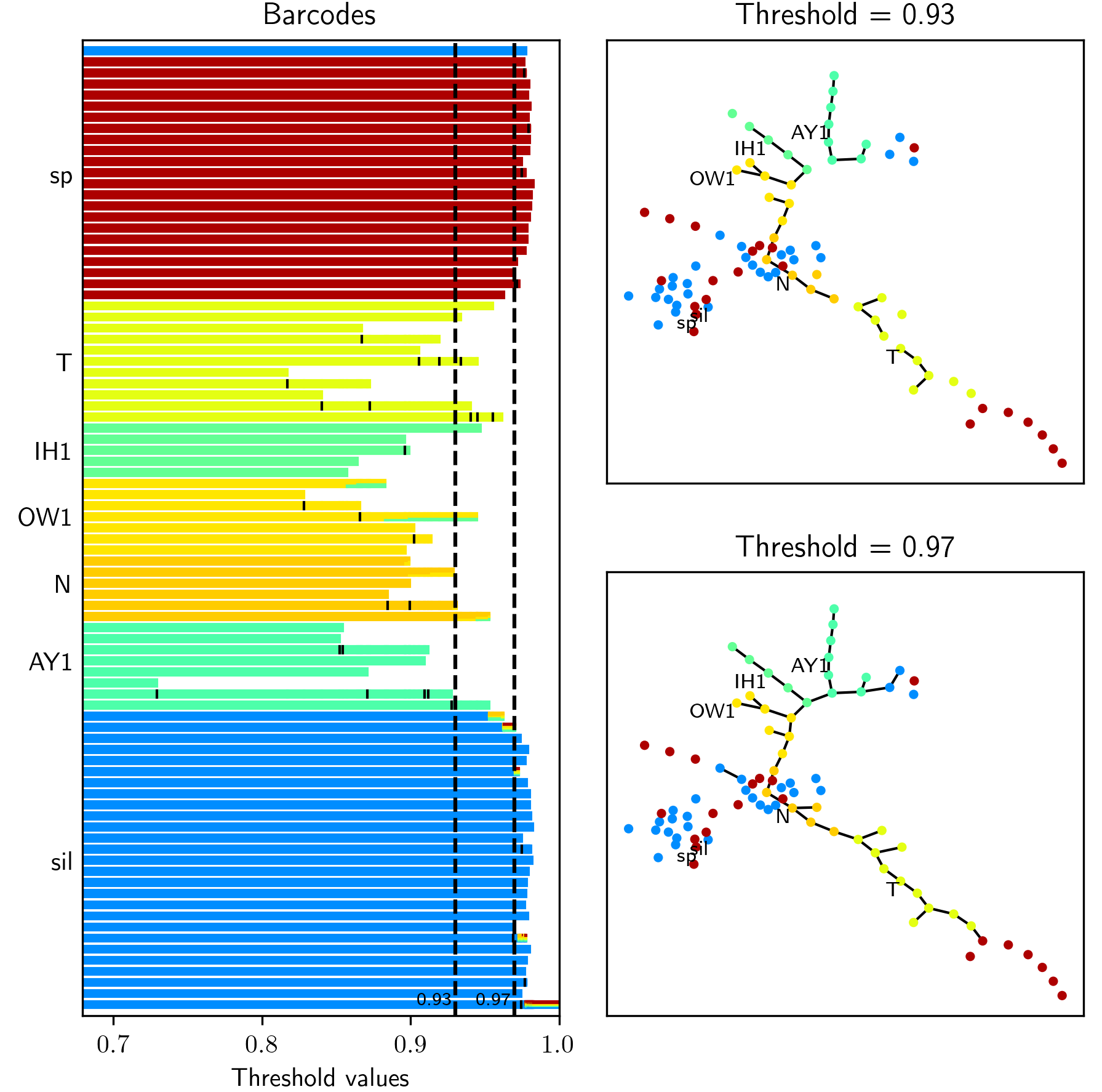}
    \vskip-.1in
\caption{Sample barcode and MST on different levels for head (2, 4) for one speech sample; sample text: ``I know it'', sample phones: ``sil AY1 N OW1 IH1 T sp''. Nodes and bars are colored with respect to the phonemes they represent. Black dashed lines show barcode levels corresponding to the trees on the right. Separators inside the bars show levels where nodes from the same phoneme are joined to the bar's component.}\label{fig:trees}
\vskip-.2in
\end{figure}




For individual model separation, we collected all samples produced by a given synthetic model and an equal amount of bonafide samples (real speech) randomly selected from train and validation sets. For each HuBERT attention head, we calculate the distributions of the $\hzsym$ feature for synthetic and real samples and rank the heads by separation quality defined as
\begin{align}
\mathrm{SQ}_{1, 2} = \frac{\vert m_1 - m_2 \vert}{\max(\sigma_1, \sigma_2)}
, 
\end{align}
where $m_i$ are the means and $\sigma_i$ are the variances of classes. Fig.~\ref{fig:sep_charts} shows sample separations; it turns out that for every voice model there are several heads that separate them well ($\mathrm{SQ}>1$) and for most models there are heads that separate them very well ($\mathrm{SQ}>3$). The best heads are different and are usually situated in the middle-to-top layers of HuBERT. The worst results were obtained on the A19 model that was specially fine-tuned on evaluation data (best $\mathrm{SQ} = 1.45$). Separation quality of the threshold classifier for the best head varies from EER $0.03\%$ for the A14 model ($\mathrm{SQ} = 3.5$) to EER $36.5\%$ for the A19 model. 
%
For individual speaker separation, a similar approach on ASVSpoof also shows that every pair of speakers has heads with good separation ($\mathrm{SQ}>1$), but this time there are no heads with $\mathrm{SQ}>3$. Fig.~\ref{fig:sep_charts}(d-e) shows sample separations. On the other hand, more general tasks are not handled well by individual heads; e.g. the best achieved separation of male vs female speakers has $\mathrm{SQ}=0.72$. 

For emotion pair separation our approach yields results that are quite similar with those for model and speaker separation. They, and additional results for other experiments from this section, are presented at the website\footnote{{\scriptsize 
\url{topohubert.github.io/speech-topology-webpages}
}}.




\noindent
\textbf{Attention maps and classical acoustic features}.
HuBERT model works with raw signals, but its attention mechanisms can extract a lot of varied information, so we searched for potential similarities between TDA features and common acoustic features extracted to process speech: mel-frequency cepstral coefficients (MFCC) and perceptual linear predictive features (PLP). Following~\cite{borzi2022synthetic}, we extract the acoustic features for samples from the ASVSpoof dataset (real human speech only) and compute Pearson correlations between them and attention features. Fig.~\ref{fig:pcc_TDA_spectral} shows the results for $\hzpc$; heads that have a strong correlation between $\hzpc$ and mean MFCC and/or mean PLP are marked in Fig.~\ref{fig:pcc_TDA_spectral}. Note that PLP features are detected mainly on lower layers, and MFCC on the middle and top layers of the model. 




\noindent
\textbf{Interpretation of 0-dimensional barcodes}.  
Fig.~\ref{fig:trees} illustrates how the $H_0$ barcode can capture hierarchical information in a speech sample.
We take a sample from \emph{LibriSpeech} with phoneme alignment~\cite{Lugosch2019} and consider the graph built from the adjacency matrix by formula~(\ref{eq:adjuc}). 



Each bar is colored with respect to the proportion of each phoneme in the corresponding connected component. When edges are absent, each bar corresponds to the component of one vertex, so the entire component belongs to a single phoneme. For larger thresholds, mixed components will appear, which would change the bar's color. 
However, on the left of Fig.~\ref{fig:trees} we can see that all bars do not change color until very high threshold values.
Moreover, there is a clear difference in length between bars corresponding to phonemes (green, yellow, orange) and bars in non-speech parts labeled as \textit{sp} (pause, red) or \textit{sil} (silence, blue). In general, the barcode reflects the hierarchy: there is a large number of short single-phoneme bars, a few longer bars connect phonemes to each other, and finally the longest bars in Fig.~\ref{fig:trees} correspond to non-speech parts of the sample.



These structure levels can be observed in the right-hand part of Fig.~\ref{fig:trees}. The top right graph shows each phoneme in a separate connected component, while the bottom right graph shows the whole phrase covered by the tree, but silence and pause nodes stay apart. 
It means that as the threshold increases, the vertices first form phoneme clusters, then connect the whole sentence, and only after that silence is attached. 
This tendency can be observed in many examples on specific heads.
\vskip-.1in


\section{Conclusion}\label{sec:concl}
In this work, we have applied topological data analysis to solving downstream tasks based on a pretrained HuBERT model and analysis and interpretation of individual attention heads. We have shown that TDA yields compact feature sets that give excellent results for tasks such as emotion recognition and speaker classification, including a new state of the art result on CREMA-D. Besides, we have shown how TDA can help interpret individual heads in a Transformer-based architecture, and shown that the structure defined by TDA features corresponds well to the semantic structure of a speech sample. We believe that topological analysis is an important and currently underexplored venue of research for large machine learning models such as Transformers, and propose TDA as a potentially fruitful direction of study.

\section{Acknowledgements}\label{sec:gracias}
The work of E.T., S.B. and E.B. was partially supported by the Analytical center under the RF Government (subsidy agreement 000000D730321P5Q0002, Grant No. 70-2021-00145 02.11.2021).

\bibliographystyle{IEEEtran}
\bibliography{mybib}

\begin{thebibliography}{10}
\providecommand{\url}[1]{#1}
\csname url@samestyle\endcsname
\providecommand{\newblock}{\relax}
\providecommand{\bibinfo}[2]{#2}
\providecommand{\BIBentrySTDinterwordspacing}{\spaceskip=0pt\relax}
\providecommand{\BIBentryALTinterwordstretchfactor}{4}
\providecommand{\BIBentryALTinterwordspacing}{\spaceskip=\fontdimen2\font plus
\BIBentryALTinterwordstretchfactor\fontdimen3\font minus
  \fontdimen4\font\relax}
\providecommand{\BIBforeignlanguage}[2]{{%
\expandafter\ifx\csname l@#1\endcsname\relax
\typeout{** WARNING: IEEEtran.bst: No hyphenation pattern has been}%
\typeout{** loaded for the language `#1'. Using the pattern for}%
\typeout{** the default language instead.}%
\else
\language=\csname l@#1\endcsname
\fi
#2}}
\providecommand{\BIBdecl}{\relax}
\BIBdecl

\bibitem{9585401}
W.-N. Hsu, B.~Bolte, Y.-H.~H. Tsai, K.~Lakhotia, R.~Salakhutdinov, and
  A.~Mohamed, ``Hubert: Self-supervised speech representation learning by
  masked prediction of hidden units,'' \emph{IEEE/ACM Transactions on Audio,
  Speech, and Language Processing}, vol.~29, pp. 3451--3460, 2021.

\bibitem{baevski2020wav2vec}
A.~Baevski, Y.~Zhou, A.~Mohamed, and M.~Auli, ``wav2vec 2.0: A framework for
  self-supervised learning of speech representations,'' \emph{Advances in
  Neural Information Processing Systems}, vol.~33, pp. 12\,449--12\,460, 2020.

\bibitem{chung2021w2v}
Y.-A. Chung, Y.~Zhang, W.~Han, C.-C. Chiu, J.~Qin, R.~Pang, and Y.~Wu,
  ``W2v-bert: Combining contrastive learning and masked language modeling for
  self-supervised speech pre-training,'' in \emph{2021 IEEE Automatic Speech
  Recognition and Understanding Workshop (ASRU)}.\hskip 1em plus 0.5em minus
  0.4em\relax IEEE, 2021, pp. 244--250.

\bibitem{Yang2021SUPERBSP}
S.~wen Yang, P.-H. Chi, Y.-S. Chuang, C.-I. Lai, K.~Lakhotia, Y.~Y. Lin, A.~T.
  Liu, J.~Shi, X.~Chang, G.-T. Lin, T.~hsien Huang, W.-C. Tseng, K.~tik Lee,
  D.-R. Liu, Z.~Huang, S.~Dong, S.-W. Li, S.~Watanabe, A.~rahman Mohamed, and
  H.~yi~Lee, ``Superb: Speech processing universal performance benchmark,'' in
  \emph{Interspeech}, 2021.

\bibitem{vulic2020probing}
I.~Vuli{\'c}, E.~M. Ponti, R.~Litschko, G.~Glava{\v{s}}, and A.~Korhonen,
  ``Probing pretrained language models for lexical semantics,'' in
  \emph{Proceedings of the 2020 Conference on Empirical Methods in Natural
  Language Processing (EMNLP)}, 2020, pp. 7222--7240.

\bibitem{kushnareva-etal-2021-artificial}
\BIBentryALTinterwordspacing
L.~Kushnareva, D.~Cherniavskii, V.~Mikhailov, E.~Artemova, S.~Barannikov,
  A.~Bernstein, I.~Piontkovskaya, D.~Piontkovski, and E.~Burnaev, ``Artificial
  text detection via examining the topology of attention maps,'' in
  \emph{Proceedings of the 2021 Conference on Empirical Methods in Natural
  Language Processing}.\hskip 1em plus 0.5em minus 0.4em\relax Online and Punta
  Cana, Dominican Republic: Association for Computational Linguistics, Nov.
  2021, pp. 635--649. [Online]. Available:
  \url{https://aclanthology.org/2021.emnlp-main.50}
\BIBentrySTDinterwordspacing

\bibitem{judgements}
\BIBentryALTinterwordspacing
D.~Cherniavskii, E.~Tulchinskii, V.~Mikhailov, I.~Proskurina, L.~Kushnareva,
  E.~Artemova, S.~Barannikov, I.~Piontkovskaya, D.~Piontkovski, and E.~Burnaev,
  ``Acceptability judgements via examining the topology of attention maps,'' in
  \emph{Findings of the Association for Computational Linguistics: EMNLP
  2022}.\hskip 1em plus 0.5em minus 0.4em\relax Abu Dhabi, United Arab
  Emirates: Association for Computational Linguistics, dec 2022, pp. 88--107.
  [Online]. Available: \url{https://aclanthology.org/2022.findings-emnlp.7}
\BIBentrySTDinterwordspacing

\bibitem{RUCCO2017130}
M.~Rucco, R.~Gonzalez-Diaz, M.-J. Jimenez, N.~Atienza, C.~Cristalli,
  E.~Concettoni, A.~Ferrante, and E.~Merelli, ``A new topological entropy-based
  approach for measuring similarities among piecewise linear functions,''
  \emph{Signal Processing}, vol. 134, pp. 130--138, 2017.

\bibitem{Gonzalez-Diaz}
R.~Gonzalez-Diaz, E.~Paluzo-Hidalgo, and J.~Quesada, ``Towards emotion
  recognition: A persistent entropy application,'' in \emph{Computational
  Topology in Image Context}.\hskip 1em plus 0.5em minus 0.4em\relax Springer
  International Publishing, 2019, pp. 96--109.

\bibitem{9747228}
T.~Fireaizen, S.~Ron, and O.~Bobrowski, ``Alarm sound detection using
  topological signal processing,'' in \emph{ICASSP 2022 - 2022 IEEE
  International Conference on Acoustics, Speech and Signal Processing
  (ICASSP)}, 2022, pp. 211--215.

\bibitem{vukovic-etal-2022-dialogue}
R.~Vukovic, M.~Heck, B.~Ruppik, C.~van Niekerk, M.~Zibrowius, and M.~Gasic,
  ``Dialogue term extraction using transfer learning and topological data
  analysis,'' in \emph{Proceedings of the 23rd Annual Meeting of the Special
  Interest Group on Discourse and Dialogue}.\hskip 1em plus 0.5em minus
  0.4em\relax Edinburgh, UK: Association for Computational Linguistics, Sep.
  2022, pp. 564--581.

\bibitem{haghighatkhah-etal-2022-story}
P.~Haghighatkhah, A.~Fokkens, P.~Sommerauer, B.~Speckmann, and K.~Verbeek,
  ``Story trees: Representing documents using topological persistence,'' in
  \emph{Proceedings of the Thirteenth Language Resources and Evaluation
  Conference}.\hskip 1em plus 0.5em minus 0.4em\relax Marseille, France:
  European Language Resources Association, Jun. 2022, pp. 2413--2429.

\bibitem{topologyofdeep}
G.~Naitzat and A.~Zhitnikov, ``Topology of deep neural networks,''
  \emph{Journal of Machine Learning Research}, vol.~21, pp. 1--40, 01 2020.

\bibitem{barannikov2021representation}
S.~Barannikov, I.~Trofimov, N.~Balabin, and E.~Burnaev, ``{Representation
  Topology Divergence: A Method for Comparing Neural Network
  Representations},'' in \emph{International Conference on Machine
  Learning}.\hskip 1em plus 0.5em minus 0.4em\relax PMLR 162, 2022.

\bibitem{barannikov2021canonical}
S.~Barannikov, ``{Canonical Forms = {P}ersistence Diagrams. {T}utorial},'' in
  \emph{European Workshop on Computational Geometry (EuroCG 2021)}, 2021.

\bibitem{10.3389/frai.2021.667963}
F.~Chazal and B.~Michel, ``An introduction to topological data analysis:
  Fundamental and practical aspects for data scientists,'' \emph{Frontiers in
  Artificial Intelligence}, vol.~4, 2021.

\bibitem{barannikov1994}
S.~Barannikov, ``The framed {M}orse complex and its invariants,''
  \emph{Advances in Soviet Mathematics}, vol.~21, pp. 93--115, 1994.

\bibitem{kenton2019bert}
J.~D. M.-W.~C. Kenton and L.~K. Toutanova, ``Bert: Pre-training of deep
  bidirectional transformers for language understanding,'' in \emph{Proceedings
  of NAACL-HLT}, 2019, pp. 4171--4186.

\bibitem{Busso2008IEMOCAPIE}
C.~Busso, M.~Bulut, C.-C. Lee, E.~A. Kazemzadeh, E.~M. Provost, S.~Kim, J.~N.
  Chang, S.~Lee, and S.~S. Narayanan, ``Iemocap: interactive emotional dyadic
  motion capture database,'' \emph{Language Resources and Evaluation}, vol.~42,
  pp. 335--359, 2008.

\bibitem{8639633}
G.~Ramet, P.~N. Garner, M.~Baeriswyl, and A.~Lazaridis, ``Context-aware
  attention mechanism for speech emotion recognition,'' in \emph{2018 IEEE
  Spoken Language Technology Workshop (SLT)}, 2018, pp. 126--131.

\bibitem{9747460}
I.~Gat, H.~Aronowitz, W.~Zhu, E.~Morais, and R.~Hoory, ``Speaker normalization
  for self-supervised speech emotion recognition,'' in \emph{ICASSP 2022 - 2022
  IEEE International Conference on Acoustics, Speech and Signal Processing
  (ICASSP)}, 2022, pp. 7342--7346.

\bibitem{6849440}
H.~Cao, D.~G. Cooper, M.~K. Keutmann, R.~C. Gur, A.~Nenkova, and R.~Verma,
  ``Crema-d: Crowd-sourced emotional multimodal actors dataset,'' \emph{IEEE
  Transactions on Affective Computing}, vol.~5, no.~4, pp. 377--390, 2014.

\bibitem{9358099}
A.~Nautsch, X.~Wang, N.~Evans, T.~H. Kinnunen, V.~Vestman, M.~Todisco,
  H.~Delgado, M.~Sahidullah, J.~Yamagishi, and K.~A. Lee, ``Asvspoof 2019:
  Spoofing countermeasures for the detection of synthesized, converted and
  replayed speech,'' \emph{IEEE Transactions on Biometrics, Behavior, and
  Identity Science}, vol.~3, no.~2, pp. 252--265, 2021.

\bibitem{Yu2018SpoofingDI}
H.~Yu, Z.~Tan, Z.~Ma, R.~Martin, and J.~Guo, ``Spoofing detection in automatic
  speaker verification systems using dnn classifiers and dynamic acoustic
  features,'' \emph{IEEE Transactions on Neural Networks and Learning Systems},
  vol.~29, pp. 4633--4644, 2018.

\bibitem{nagrani17_interspeech}
A.~Nagrani, J.~S. Chung, and A.~Zisserman, ``{VoxCeleb: A Large-Scale Speaker
  Identification Dataset},'' in \emph{Proc. Interspeech 2017}, 2017, pp.
  2616--2620.

\bibitem{7178964}
V.~Panayotov, G.~Chen, D.~Povey, and S.~Khudanpur, ``Librispeech: An asr corpus
  based on public domain audio books,'' in \emph{2015 IEEE International
  Conference on Acoustics, Speech and Signal Processing (ICASSP)}, 2015, pp.
  5206--5210.

\bibitem{ristea22_interspeech}
N.~C. Ristea, R.~T. Ionescu, and F.~S. Khan, ``{SepTr: Separable Transformer
  for Audio Spectrogram Processing},'' in \emph{Proc. Interspeech 2022}, 2022,
  pp. 4103--4107.

\bibitem{borzi2022synthetic}
S.~Borz{\`\i}, O.~Giudice, F.~Stanco, and D.~Allegra, ``Is synthetic voice
  detection going into the right direction?'' in \emph{Proceedings of the
  IEEE/CVF Conference on Computer Vision and Pattern Recognition}, 2022, pp.
  71--80.

\bibitem{Lugosch2019}
\BIBentryALTinterwordspacing
L.~Lugosch, M.~Ravanelli, P.~Ignoto, V.~S. Tomar, and Y.~Bengio, ``{Speech
  Model Pre-Training for End-to-End Spoken Language Understanding},'' in
  \emph{Proc. Interspeech 2019}, 2019, pp. 814--818. [Online]. Available:
  \url{http://dx.doi.org/10.21437/Interspeech.2019-2396}
\BIBentrySTDinterwordspacing

\end{thebibliography}

\end{document}